\documentclass[aps,prb,twocolumn,superscriptaddress,showpacs,showkeys,preprint,onecolumn]{revtex4-2}

\usepackage{graphicx}
\usepackage{amsmath}
\usepackage{color}

\begin{document}

\title{Observation of the Tetragonal-Orthorhombic transition in polycrystalline YBa$_2$Cu$_3$O$_{6+\delta}$ samples during the oxidation process at constant temperature via thermogravimetry and differential thermometry analysis}

\author{Lorenzo Gallo}
\address{Universidad Nacional de Rosario, Pellegrini 1120, S2000EZP Rosario, Santa F\'e, Argentina.}
\author{Cesar E. Sobrero}
\address{Universidad Nacional de Rosario, Pellegrini 1120, S2000EZP Rosario, Santa F\'e, Argentina.}
\address{Institute of Physics Rosario, CONICET-UNR, Bv. 27 de Febrero 210bis, S2000EZP Rosario, Santa F\'e, Argentina.}
\author{Jorge A. Malarría}
\address{Universidad Nacional de Rosario, Pellegrini 1120, S2000EZP Rosario, Santa F\'e, Argentina.}
\address{Institute of Physics Rosario, CONICET-UNR, Bv. 27 de Febrero 210bis, S2000EZP Rosario, Santa F\'e, Argentina.}
\author{Roberto F. Luccas}
\address{Institute of Physics Rosario, CONICET-UNR, Bv. 27 de Febrero 210bis, S2000EZP Rosario, Santa F\'e, Argentina.}

\date{\today}

\begin{abstract}
This work examines the oxygenation process of polycrystalline powder of YBa$_2$Cu$_3$O$_{6+\delta}$ through constant temperature thermogravimetric analysis.
Starting from completely deoxygenated samples ($\delta$ = 0), both the mass evolution in an oxygen atmosphere and a differential temperature value referenced to an inert sample (alumina powder) subjected to the same experimental conditions are recorded.
The oxygenation process itself is identified as an exothermic process accompanied by a second also exothermic associated process.
This second process is identified as the Tetragonal-Orthorhombic transition present in the material.
Based on the obtained results, the structural phase diagram of the material is reconstructed.
Issues of metastability of the structural phases, as well as the potential of the equipment used to obtain these results comparable to those obtained in large facilities, are discussed.
Reinterpretations of previously discussed works in the literature are inferred.
\end{abstract}

\maketitle

\section{Introduction}

The superconductor material YBa$_2$Cu$_3$O$_{6+\delta}$ (YBCO) in its deoxygenated state, $\delta$ = 0, appears in a Tetragonal phase with identical a and b unit cell parameters\cite{Waldram96}.
Its typical triple perovskite-like structure is completed with a c parameter value approximately 3 times the a parameter value.
This observed symmetry upon 90-degree rotations on the c direction, is broken when the material is oxygenated (where 0 $\le$ $\delta$ $\le$ 1), leading to a transition to an Orthorhombic phase\cite{Gallagher87,Eatough87}.

During oxygenation, the oxygen atoms gained by YBCO are located at the apical sites of the unit cell, at any of the 4 sites the cell has both at its base and its top (base of the next unit cell)\cite{David87}.
This randomness in the location of absorbed oxygen allows YBCO to maintain its Tetragonal structure at low degrees of oxygenation.
As the material oxygenates and the oxygen concentration $\delta$ increases, the O atoms begin to order, forming Cu-O chains.
This results in a preference in the location of oxygen gained by the material, taking the b directions in the a-b plane of the YBCO unit cell.
The breaking of this symmetry leads to an expansion of the unit cell in the b direction, resulting in the Tetragonal-Orthorhombic (T-O) transition known in the material\cite{OBryanGallagher87PartII}.

Initially, and considering only the positions that the apical oxygens take in the unit cell, the material undergoes this T-O transition from a disordered phase to an ordered phase with equal probability of occupation of all apical oxygen sites in the direction of the b parameter of the unit cell.
This order in the Orthorhombic phase is called Ortho-I superstructure; and the T-O transition, known to occur from $\delta$ values $\approx$ 0.5, is assumed to this phase with incomplete filling of the available oxygen sites\cite{Ceder90,Andersen90,Zimmermann03,Andersen99}.
However, it has been measured by high-energy X-ray diffraction experiments\cite{Zimmermann03} that there are different types of superstructures, leading to Orthorhombic phases called Ortho-I, Ortho-II, Ortho-III, Ortho-V, and Ortho-VIII.
Among them, the Ortho-I superstructure predominates at both high oxygenation temperatures and high $\delta$ values, coinciding with being the only one with long-range interaction.
Of the remaining superstructures, the one called Ortho-II predominates at low oxygenation temperatures and low $\delta$ values, coinciding with being the only remaining one with 3D order (like Ortho-I).
The other superstructures, found at low oxygenation temperatures and at $\delta$ values that place them between Ortho-II and Ortho-I phases, are all 2D ordered\cite{Zimmermann03}

In the work presented here, we study the evolution of the mass of YBCO powder as a function of oxygenation time t$_{Oxyg}$ and for different Oxygenation Temperatures T$_O$ values, in the range from 320$^o$C to 790$^o$C.
We work with YBCO powder obtained by grinding superconducting YBCO pellets.
The mass variation results during the oxygenation process were obtained with a commercial thermo-gravimetric analysis (TGA) equipment, which also takes data from differential thermometry (DTA) in contrast with an inert sample (alumina powder).
The results of DTA show the oxygenation of the material as an exothermic process that is accompanied by a second exothermic process associated with the T-O transition present in the material.
In conjunction with the results of TGA, we observe not only the T-O transition to Ortho-I phase at high T$_O$ values and to Ortho-II phase at low T$_O$ values, but also the interface between Ortho-I and Ortho-II superstructures in the Orthorhombic phase, which depends on the material's oxygenation temperature.
We attribute the observation of the interface between superstructures in our experiments at constant T$_O$, to both the type of material used and the method employed to conduct our experiments.
Both allowing the Tetragonal phase to evolve metastably along this interface.
The metastability conditions are discussed in the work.
Moreover, the outstanding potential of the equipment used to obtain results comparable to those obtained in large facilities is also discussed.
Additionally, different interpretations of results regarding the theme of oxygenation in the bibliography are proposed.

\section{Experimental details}

We worked with YBCO powder.
A PANalytical Empyrean X-ray diffraction system was used.
X-ray diffraction analysis confirmed that YBCO powder was in either Tetragonal or Orthorhombic phase depending on the material's degree of oxygenation.
An Olympus PME 3 microscope was used, processing the images obtained with TSView software, to determine the grain size of the YBCO powder.
Approximately 4.5 g of powder was obtained with a grain size of 5 $\mu$m (see Supplementary Materials).
A Shimadzu DTG-60H was used for TGA and DTA studies.
Pt crucibles and alumina powder as reference sample were used for each measurement.
Two different gases were used for both TGA and DTA studies: Nitrogen as an inert gas, and Oxygen.
Approximately 55 mg of YBCO powder was used for measurements made in the TGA keeping the error below the 4\%, always using powder not previously measured.

For X-ray measurements, YBCO powder was used under the same oxygenation conditions as observed in TGA experiments.
Three different oxygenation levels were studied: completely deoxygenated ($\delta$ = 0), oxygenated at low T$_O$ (350 $^o$C, $\delta$ $\approx$ 0.93(5)), and oxygenated at high T$_O$ (689 $^o$C, $\delta$ $\approx$ 0.77(5)).

\section{Results}

\begin{figure}
\begin{center}
\includegraphics[width=0.9 \columnwidth]{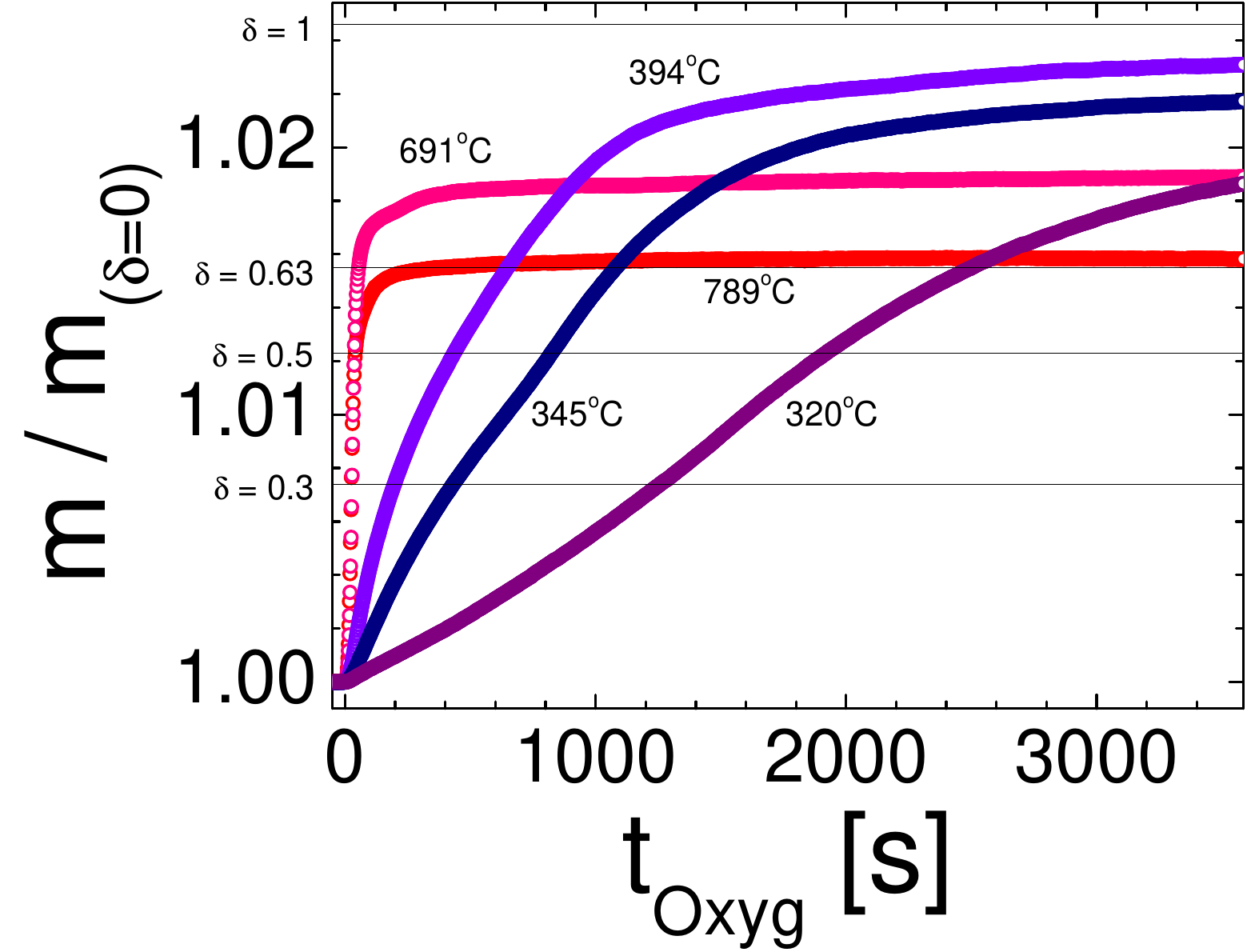}
\caption{\label{Fig1}
Typical evolution at constant T$_O$ of relative mass with respect to t$_{Oxyg}$, for different values of T$_O$ ranging from 320 $^o$C to 789 $^o$C.
Here, m represents the measured mass and m$_{(\delta = 0)}$ the mass of the material when fully deoxygenated.
Various oxygenation values $\delta$ are indicated as references on the m/m$_{(\delta = 0)}$ scale ($\delta$ = 1, $\delta$ = 0.63, $\delta$ = 0.5 and $\delta$ = 0.3).
A growth in mass as a function of t$_{Oxyg}$ is observed.
Additionally, both the rate of mass growth and the final saturated value depend on the T$_O$ value.
We observed that for T$_O$ values below 450 $^o$C the system continues to evolve within the range of shown t$_{Oxyg}$ (t$_{Oxyg}$ $<$ 1 hr).}
\end{center}
\end{figure}

The Fig. \ref{Fig1} provides typical evolution of relative masses m/m$_{(\delta = 0)}$ as a function of t$_{Oxyg}$ in a TGA experiment for selected T$_O$ values from the range under study.
On it, m represents the sample mass, and m$_{(\delta = 0)}$ represents the mass of the fully deoxygenated sample.
The theoretical mass value corresponding to the maximum possible Oxygenation, $\delta$ = 1, is marked as a reference on the graph.
Other $\delta$ values are also marked as references: $\delta$ = 0.3, $\delta$ = 0.5, and $\delta$ = 0.63\cite{Ceder90,Zimmermann03,Gallagher87,OBryanGallagher87PartII,Gallagher87PartI}.
The sample mass clearly increases as a function of exposure time in an Oxygen-saturated atmosphere, which we directly associate with the material's degree of Oxygenation.
The rate at which the material evolves its degree of Oxygenation, as well as the saturation value it reaches, clearly depend of T$_O$.

More over, three samples with different degrees of Oxygenation were studied by X-ray experiments.
One sample under conditions of $\delta$ = 0 (t$_{Oxyg}$ = 0), and two samples Oxygenated for 1 hour at T$_O$ = 350 $^o$C ($\delta$ $\approx$ 0.93(5)) and at T$_O$ = 689 $^o$C ($\delta$ $\approx$ 0.77(5)).
For the deoxygenated sample, a Tetragonal structure was observed, while for both Oxygenated samples, an Orthorhombic structure was observed.
This corresponds to literature data, where the structural T-O transition is expected to occur at $\delta$ values beyond 0.63 (T-O in the $\delta$ range 0.5 - 0.63)\cite{Gallagher87}.

\begin{figure}
\begin{center}
\includegraphics[width=0.9 \columnwidth]{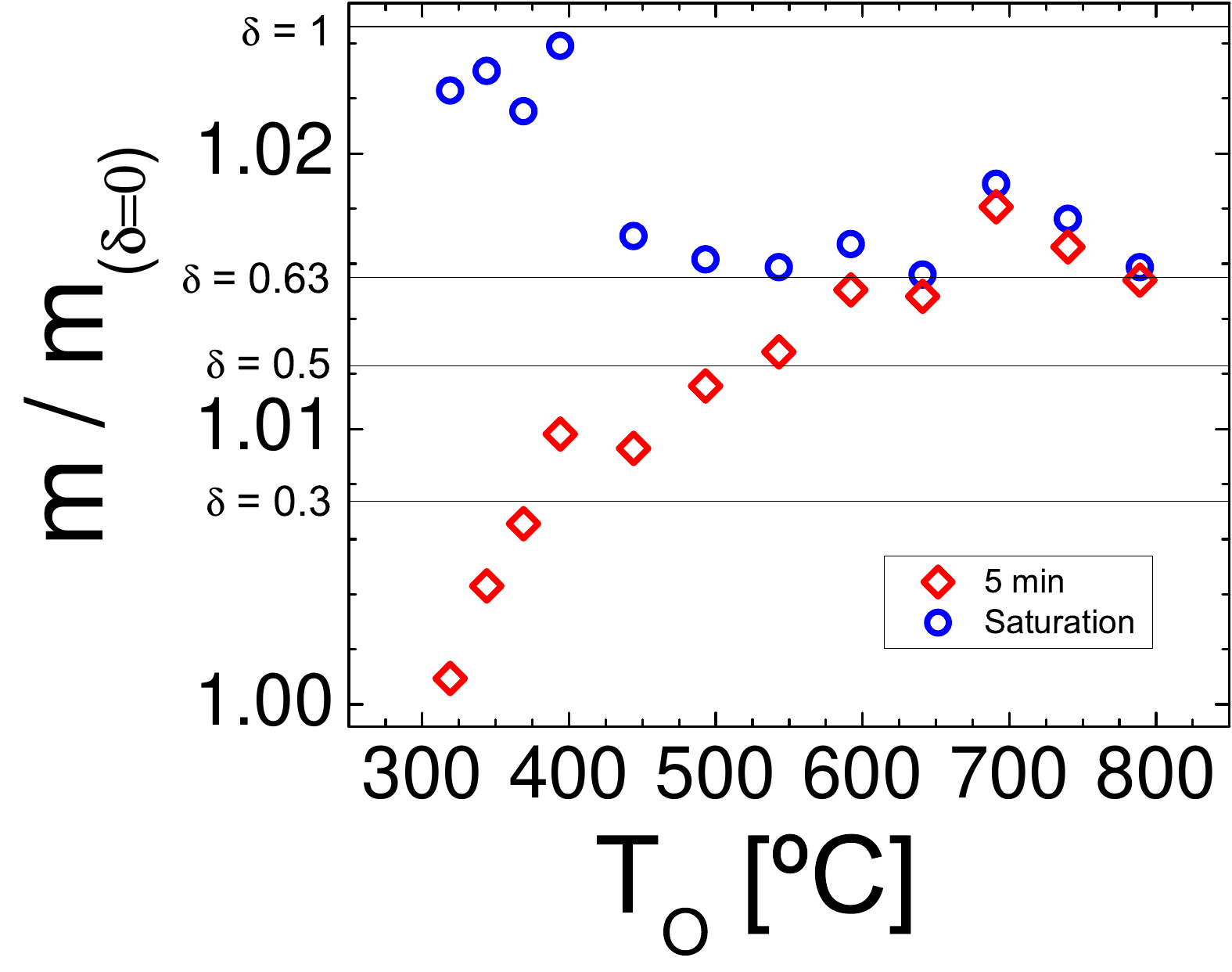}
\caption{\label{Fig2}
Measured values in oxygenation experiments at constant T$_O$ of relative mass m/m$_{(\delta = 0)}$ for two specific times, as a function of T$_O$.
Values are shown for t$_{Oxyg}$ = 5 min (red diamonds) and at saturation condition for t$_{Oxyg}$ $\ge$ 1 hs (blue circles).
It is observed that, despite obtaining higher oxygenation values at low temperatures, above 600 $^o$C an oxygenation value of the material near to the saturation is reached in very short times.}
\end{center}
\end{figure}

Observing the saturation times in TGA data, we make an initial comparison of the Oxygenation values reached in short and long times.
We arbitrarily take 5 minutes vs. saturation time as references, respectively.
The Fig. \ref{Fig2} provides the value reached by the relative mass increase m/m$_{(\delta = 0)}$ at these two conditions for each of the T$_O$ values studied.
The maximum Oxygenation value at saturation is observed for T$_O$ = 394 $^o$C and corresponds to a $\delta$ value of 0.97.
Then, for T$_O$ above 600 $^o$C, we note that the difference between the Oxygenation values reached at t$_{Oxyg}$ = 5 minutes and for saturation values is minimal (below the 5\%), observing a maximum in Oxygenation for T$_O$ = 691 $^o$C.

\begin{figure}
\begin{center}
\includegraphics[width=0.6 \columnwidth]{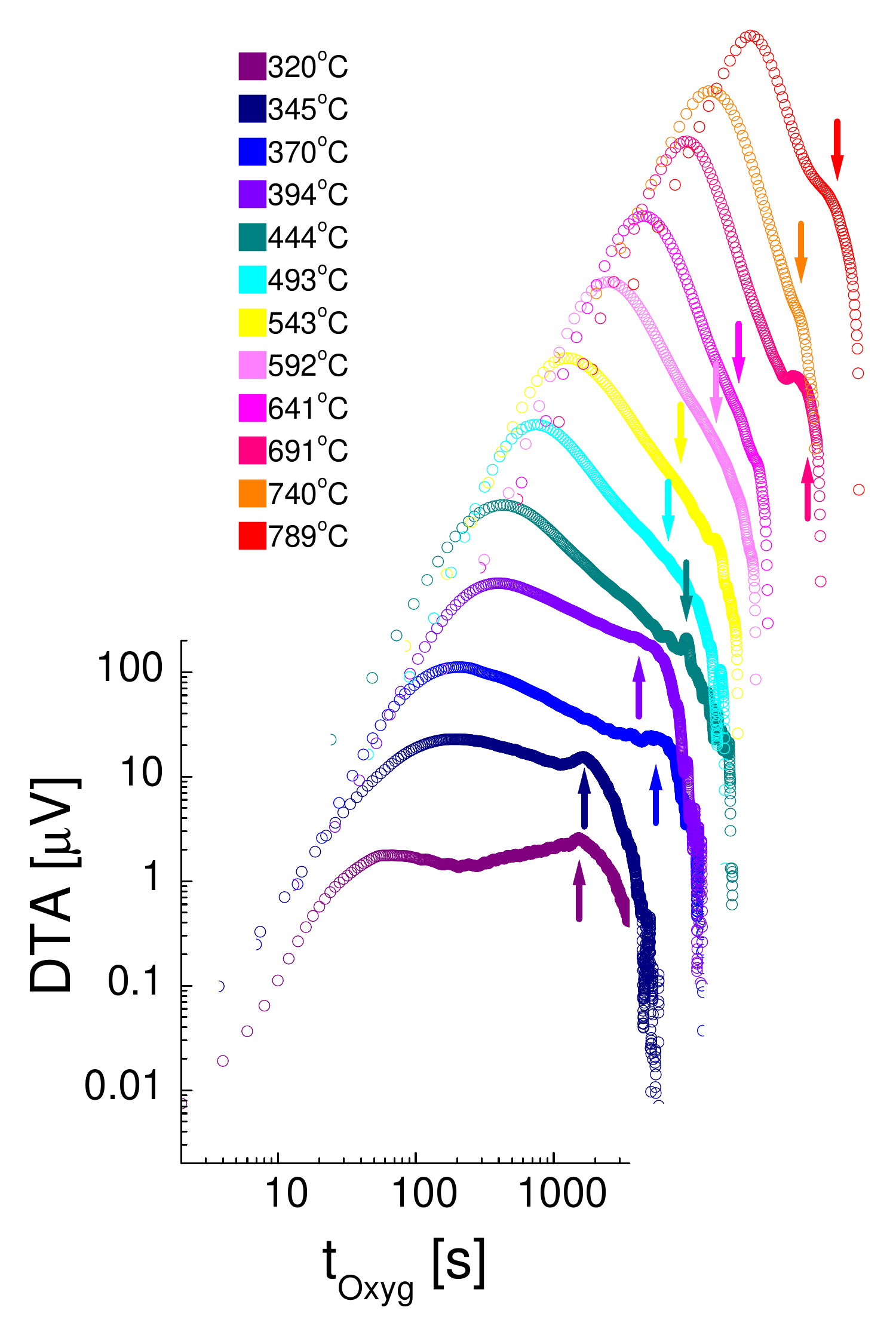}
\caption{\label{Fig3}
DTA signal of the material during the oxygenation process at constant T$_O$, as a function of t$_{Oxyg}$ and for the studied values of T$_O$ (in the range from 320 $^o$C to 789 $^o$C.
The temperature during the oxygenation process of an inert sample (alumina powder) is taken as a reference.
The data is shown on a double logarithmic scale.
Data for the lowest T$_O$ value corresponds to the axis pair shown on the graph.
For a better observation, data corresponds to any other T$_O$ value has been shifted a fixed value upwards and to the right relative to the immediately preceding T$_O$ value data set.
Two exothermic processes are observed during the material oxygenation.
The first, predominant at high temperatures, corresponds to the oxygenation of the material.
The second, indicated by arrows in the curves, is associated with the T-O transition present in the material.}
\end{center}
\end{figure}

On the other hand we observe the DTA results yielded by the experiment.
The Fig. \ref{Fig3} shows the DTA value as a function of t$_{Oxyg}$ for each of the T$_O$ values studied.
The data are shown on a double logarithmic scale where, for a better observation, for each T$_O$ the graph has been shifted a fixed value upwards and to the right relative to the immediately preceding T$_O$ value, keeping data for T$_O$ = 320 $^o$C with no shifting to the axis pair shown on the graph.
It is observed in all curves that the first peak, which above T$_O$ = 450 $^o$C becomes the predominant effect, starts with oxygenation process and ends at oxygen saturation of the material.
Therefore, we associated this peak with YBCO oxygen absorption process, showing it as a clearly exothermic one.
Furthermore, we can observe a second peak that for high T$_O$ is practically drawn as a shoulder in the curves; while, for low T$_O$ values, it takes on the relevance of the first peak corresponding to the Oxygenation process.
This second peak is indicated in the graphic with arrows, and we associate it with an imprint of the T-O transition present in the material.

\begin{figure}
\begin{center}
\includegraphics[width=0.9 \columnwidth]{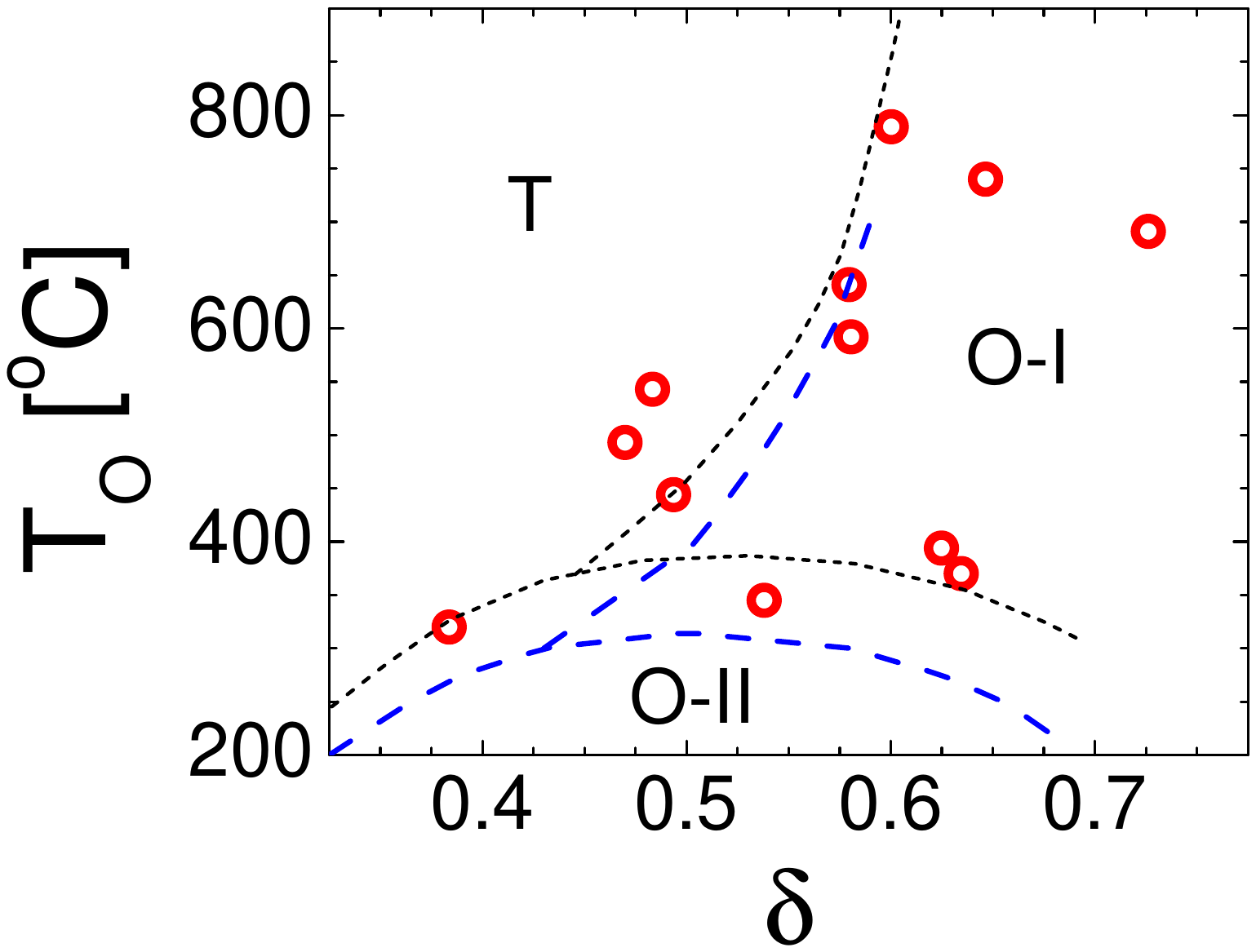}
\caption{\label{Fig4}
Structural phase diagram of YBCO in a T$_O$ vs. $\delta$ graph, obtained from the position of the T-O transition observed in experiments of DTA for the oxygenation process at constant T$_O$.
Measured data are shown as red circles.
Values for the theoretically predicted structural phase diagram according to ref. \citep{Ceder90} are shown as blue dashed lines.
A shift towards higher temperatures of the measured data in this work compared to theoretical predictions in the literature is observed.
This shift is indicated by black dotted lines as a guide to the eye.}
\end{center}
\end{figure}

Introducing the t$_{Oxyg}$ value associated to the second peak that is obtained from Fig. \ref{Fig3} in the TGA mass evolution graphic (Fig. \ref{Fig1}), we obtain a $\delta$ value associated to the T-O transition.
The Fig. \ref{Fig4} shows this result in red circles.
On it, T denotes the Tetragonal phase, O-I denotes the Ortho-I superstructure, and O-II denotes the Ortho-II superstructure.
As a reference, dashed blue lines show the values of the T-O transition proposed from first-principles calculations obtained using a 2D-ASYNNNI model\cite{Ceder90}.
The black dotted lines are only drawn as a guide to the eye.

It is observed that, in principle, the measured values at low T$_O$ are in good agreement with the theoretical line separating the O-II superstructure from the T phase and the O-I superstructure.
However, we note that these data are located at temperatures slightly above the theoretical predictions\cite{Andersen90,Zimmermann03}.
For temperatures greater than 400 $^o$C, the data measured in this work agree with the theoretical prediction for the T-O transition (O-I superstructure)\cite{Andersen90,Zimmermann03}.

\section{Discussion}

Regarding the diffusion rate of O$_2$ in the material, we noticed higher efficiency in the oxidation process for T$_O$ values above 600 $^o$C; where, for practical purposes, an oxidation experiment with t$_{Oxyg}$ = 5 min yields the same results as an experiment with t$_{Oxyg}$ = 1 hr, within 5 \% of tolerance.
Particularly, we measured optimal oxidation under t$_{Oxyg}$ minimization conditions for T$_O$ = 691 $^o$C.

Concerning the signature left by the T-O transition in TGA and DTA measurements, its appearance is clear and it corresponds to an exothermic transition.
On the other hand, a subtle change in concavity is observed in curves of relative mass as a function of t$_{Oxyg}$ at constant T$_O$ (Fig. \ref{Fig1}) at the moment of transitioning to the Orthorhombic phase.
This indicates an increase in the rate of O$_2$ absorption by the material due to the presence of the T-O transition.
Thus, in coherence with the previous assertion, we assume that the material's T-O transition contributes to the Oxygen diffusion on it.

This topic has been widely discussed in bulk samples\cite{Diko08,Diko07}.
Cracks in the material have been identified with an associated change in the Oxygenation dynamic due to their presence.
It is logical to assume that the appearance of a mechanism (like cracks) that directly facilitates O$_2$ diffusion in the material along the c axis contributes on its diffusion.
However, the cause of the appearance of such cracks is not considered in literature.
Our results show an acceleration in O$_2$ absorption associated with the T-O transition.
We can say that, for bulk samples, on the material surface and at least up to a depth of 2.5 $\mu$m (a radius associated with the grain size of our material), our results demonstrate the presence of this phenomenon.
Furthermore, the change in the lattice parameters a and b associated with the T-O transition produces an abrupt deformation in the a-b plane of the crystal structure, elongating the system in the b direction and shortening it in the a direction\cite{Gallagher87}.
In bulk samples, this abrupt deformation of the unit cell, if not accompanied by the entire system (at least 7 orders of magnitude larger than the dimensions of the unit cell for laboratory bulk samples\cite{Diko08,Diko07}), could cause the appearance of cracks.
This, coupled with the significant structural difference that can occur with very small variations of $\delta$\cite{Beyers89}, implies the need for high elasticity in the structure to prevent crack formation.
This required structural elasticity, evidently, is not present in bulk single crystal samples\cite{Diko08,Diko07}.
Therefore, our results demonstrate that it is not the appearance of cracks directly responsible for the change in the material's Oxygenation dynamics, but rather the intrinsic T-O transition of the material that works as a first addenda to the oxygenation process. Moreover, it leads to a sharp jump in the values of the unit cell parameters and consequently to the appearance of these so-called cracks on the material.

Finally, based on the results of $\delta$ associated to the T-O transition for each T$_O$ value under study, we draw the structural phase diagram of YBCO on a T$_O$ vs. $\delta$ graph (Fig. \ref{Fig4}).
Our data delineate the T-O transition and the interface between the O-I and O-II superstructures.
Initially, our results are in good agreement with predictions made using an Ising model with first-principles calculations (a 2D ASYNNNI model\cite{Ceder90}) and agree with data obtained by measuring YBCO samples via neutron powder diffraction\cite{Andersen90}.
What stands out from the data measured in this work is that they delineate the boundary between the O-I and O-II superstructures.
In previous experiments\cite{Ceder90,Andersen90,Zimmermann03}, the interface between this two superstructures has been observed only through experiments at constant $\delta$, that is, with temperature variation.
In our experiments, we assume that the quantity of defects generated in the material's crystal structure when obtaining the YBCO powder allows the Tetragonal phase dabbles metastably to much higher $\delta$ values.
We trust this only occurs at the interface between the O-I and O-II superstructures because the latter is a short-range order superstructure\cite{Zimmermann03} allowing the disordered coexistence of many O-II nucleation centers, maintaining a Tetragonal structure in the material until the effective appearance of the Orthorhombic phase with long-range order (O-I)\cite{Zimmermann03}.
We thus propose that for our experiments, the trigger for the T-O transition is the exit from the O-II dome, transitioning to the O-I superstructure, which has long-range order.

Moreover, we consider intrinsic defects in our samples are playing an important role on this structural diagram.
Frello \emph{et al.}\cite{FrelloPhD,Zimmermann03} performed high-energy X-ray diffraction experiments at low T$_O$ values (below 200 $^o$C), observing the interface between superstructures at these temperatures, i.e., 200 $^o$C below what is predicted by theoretical models\cite{Ceder90,Zimmermann03}.
In a subsequent study, by performing calculations that also consider 3D interaction terms as an extension of the 2D ASYNNNI model\cite{Andersen99}, they justify the appearance of the interface between O-I and O-II superstructures for T$_O$ around 100 $^o$C, claiming very high purity of their samples.
In our case, the method used to obtain the YBCO powder to be measured, mechanical milling using an agate mortar, can most likely introduce defects such as dislocations, inclusion planes, or any other type of structural defects into the material.
This could even explain the appearance of the roof of the O-II superstructure approximately 50 $^o$C above what is theoretically proposed\cite{Ceder90}.

On the other hand, at high T$_O$ values conditions are others.
For temperatures above 400 $^o$C, the Orthorhombic phase is purely long-range order (O-I), with no presence of other superstructures (O-III, O-V, O-VIII or even O-II) since they exist at much lower temperature values\cite{Zimmermann03}.
However, it is also noteworthy that above 600 $^o$C, relative $\delta$ values associated to the T-O transition (Fig. \ref{Fig4}) are in coherence with relative $\delta$ values associated to the Oxygen saturation of the material (Fig. \ref{Fig2}).
Additionally, rates of Oxygen saturation above 600 $^o$C are not only the highest but also similar between them (Fig. \ref{Fig2}), with very short times of saturation.
This indicates that metastability at high T$_O$ values depends not only on defects but also on the abrupt rates at which oxidation occurs (abrupt even for structural changes).
Considering this possibility, we propose that discrepancies between our data and the theoretically predicted ones for the T$_O$ transition values above 600 $^o$C, occurs due to the violent oxygenation rate that the material is exposed, incursing into the O-I superstructure in a metastable condition.

\section{Conclusions}

From YBCO powder obtained by mechanical milling of superconducting pellets, both mass evolution and differential temperature referenced to an inert sample were studied as functions of the oxidation time.
Starting from completely deoxygenated samples, a maximum oxidation of $\delta$ = 0.97 was achieved for T$_O$ = 394 $^o$C.
It was also observed that, for short times, t$_{Oxyg}$ = 5 min, the material reaches maximum oxidation for T$_O$ = 691 $^o$C, saturating within 5 \% of tolerance in the m value.
Additionally, the values of differential temperature indicated the oxidation process as an exothermic type.
It was also observed that, associated with this process, there is a second exothermic process very well differentiated, which we link to the T-O transition in the material.
Subsequently, a $\delta$ value was associated with the T-O transition for each T$_O$.
With these values, a structural phase diagram for the material was constructed, observing not only the T-O phase transition but also the interface between O-I and O-II superstructures in the Orthorhombic phase.
Finally, based on variations of our results compared to theoretical models predicting the structural phase diagram of the material\cite{Ceder90,Zimmermann03}, different mechanisms were proposed that generate metastability during the T-O transition in the oxidation process.
On one hand, at low temperatures, a mechanism of multiple nucleation centers of a superstructure with short-range order (O-II) was proposed to maintain the Tetragonal phase metastably into the Orthorhombic one.
On the other hand, at high temperatures, a mechanism where the high speed in the oxidation process plays a very important role was proposed.
Both cases are accompanied by a significant density of structural defects intrinsic to the nature of the samples with which we work.
In addition to this, the enormous potential of a thermogravimetry setup combined with \emph{in-situ} differential thermal analysis was highlighted.
With these simple and accessible results, it was possible to replicate results previously obtained in large facilities using techniques such as neutron diffraction and high-energy X-ray diffraction\cite{FrelloPhD}.

\section{Acknowledgments}
This work was partially supported by MINCyT through the PICT-2017-2898 FONCyT project.
LG performed the TGA-DTA experiments and oxygenation analysis.
CES performed the YBCO pellet growth and X-ray experiments and analysis.
JAM contributed to the YBCO pellet growth, TGA-DTA experiments and X-ray experiments.
RFL performed the TGA-DTA experiments and structural transition analysis, conducted the research and wrote out the manuscript.
All authors contributed equally to discussion of ideas and correction of the manuscript.

\bibliographystyle{elsarticle-num-names}
\bibliography{LGalloYbcoStructuralTransitionb.bib}

\end{document}